# Enhancement in superconducting transition temperature and upper critical field of $LaO_{0.8}F_{0.2}FeAs$ with antimony doping

S. J. Singh[a], J. Prakash[b], S. Patnaik[a] and A. K. Ganguli[b]*

[a]School of Physical Sciences, Jawaharlal Nehru University, New Delhi 110067 India

[b]Department of Chemistry, Indian Institute of Technology, New Delhi 110016 India

---

*Author for correspondence

Email:- ashok@chemistry.iitd.ernet.in

Tel No. 91-11-26591511

Fax  91-11-26854715

**Abstract**

We report the synthesis and characterization of antimony doped oxypnictide superconductor, $LaO_{0.8}F_{0.2}FeAs_{1-x}Sb_x$ (x = 0.05 and 0.10). The parent compound LaOFeAs with fluorine doping exhibits superconductivity at maximum transition temperature ~ 28.5 K [11]. Here we partially substitute As by Sb ($LaO_{0.8}F_{0.2}FeAs_{1-x}Sb_x$) and observe enhancement of the transition temperature to 30.1 K. This is the only instance so far where $T_c$ increases with doping in the conducting layer (FeAs) and this leads to the highest transition temperature in any La-based oxypnictide. XRD and EDAX measurements confirm phase purity of the samples and the presence of Sb. The magneto-resistance measurements show that the value of upper critical field $H_{c2}(0)$ to be about 73 T corresponding to a coherence length ($\xi_{GL}$) of 22 Å. The Seebeck coefficient measurements indicate electron transport with strong contribution from electron-electron correlation. These results provide interesting insight to the origin of superconductivity in these novel series of compounds.

## Introduction

In a recent discovery, an iron arsenic based layered compound LaOFeAs was reported to show superconductivity on doping fluorine at oxygen sites with a transition temperature ($T_c$) of about 26 K [1]. The parent compound (LaOFeAs) crystallizes in a tetragonal layered structure similar to ZrCuSiAs type structure [2] and is made up of alternating LaO and FeAs layers stacked along the c-axis. It is now established that the FeAs layers provide the conducting pathway and LaO layers act as charge reservoirs [3]. The parent compound undergoes a first order structural transition from a tetragonal to an orthorhombic structure upon cooling below 155-160 K. Carrier doping in these compounds leads to appearance of superconductivity by suppressing the magnetic order and structural phase transition. Following the initial work, many other related compounds with $T_c$ as high as 55 K have been synthesized by replacing La with other rare earth elements such as Ce, Sm, Nd, Pr and Gd [4-8]. Recent experimental and theoretical studies indicate that these materials appear to be unconventional multiband superconductors [9-10]. It appears that the electron-phonon interaction is not strong enough to give rise to such high transition temperature, while strong ferromagnetic and antiferromagnetic fluctuations are responsible for superconductivity. We have previously reported a new methodology of synthesizing these fluorine-doped oxypnictides with the commonly available KF as a source of fluorine instead of the expensive $LaF_3$ as used in most reports [11]. We also reported superconductivity at 28.5 K for La(O/F)FeAs which is the highest $T_c$ reported so far in La based oxypnictides[11]. In this communication we report the synthesis of new Sb-doped superconductors, $LaO_{0.8}F_{0.2}FeAs_{1-x}Sb_x$, (x = 0.05 and 0.10) which show an enhancement of the Tc to 30.1 K We report the resistivity, magneto-resistance, penetration depth and Seeback coefficient studies in these new Sb-doped superconductors and discuss the possible reasons for the increase in $T_c$ with Sb doping. This is the highest $T_c$ in the family of La-based oxypnictide superconductors. Furthermore, it is now well established that carrier doping can be introduced by introducing magnetic dopant (e.g. Co) in place of Fe [12] in the conducting layer, in contrast to copper oxide based high $T_c$ materials. Here we elucidate the possibility of simultaneous enhancement in both transition temperature and upper critical field by dilute substitution of As by Sb in the conducting FeAs layer.

**Experimental**

Polycrystalline samples with nominal composition of $LaO_{0.8}F_{0.2}FeAs_{1-x}Sb_x$ with x= 0.05 and x = 0.10 were synthesized by a two step solid state reaction method by using high purity $La_2O_3$, $LaF_3$, La, FeAs, $FeSb_2$ and Fe as starting materials. FeAs was obtained by reacting Fe chips and As powder at 950 C for 24 hours. The raw materials (all with purities better than 99.9 %) were weighed according to the stoichiometric ratio in a $N_2$-filled glove- box and then sealed in evacuated silica ampoules ($10^{-4}$ torr) and heated at 950 C for 48 hours. The powder was compacted into pellets. The pellets were wrapped in Ta foil, sealed in evacuated silica ampoules and annealed at 1150 C for 48 hours and then cooled to room temperature. The phase purity was checked by powder x-ray diffraction using Cu-Kα radiation followed by energy dispersive x-ray spectroscopy.

The resistivity measurements were carried out using a cryogenic 8 T cryogen-free magnet in conjunction with a variable temperature insert (VTI). Standard four probe technique was used for transport measurements. Contacts were made using 44 gauge copper wires with conducting silver paste. The external magnetic field (0-5 T) was applied perpendicular to the probe current direction and the data were recorded during the warming cycle with heating rate of 1 K/min. Thermopower was measured in bridge geometry and the magnitude was calibrated with OFHC copper. The inductive part (real part) of the magnetic susceptibility was measured using a tunnel diode based rf penetration depth technique [13] attached to this cryogen-free magnet system. A change in magnetic state of the sample results in a change in the inductance of the coil and is reflected as a shift in the oscillator frequency which is measured by an Agilent 53131A counter. Energy dispersive analysis by X- rays (EDAX) was carried out on sintered pellets of the compounds on a Zeiss electron microscope in conjunction with a BRUKER EDX system.

**Results and discussion:**

Fig.1 shows the XRD patterns for the samples with nominal compositions of $LaO_{0.8}F_{0.2}FeAs_{0.95}Sb_{0.05}$ and $LaO_{0.8}F_{0.2}FeAs_{0.9}Sb_{0.1}$. The pattern could be indexed on a tetragonal cell with lattice parameters of **a** = 4.02017(6) Å , **c** = 8.701(2) Å for x = 0.05 , **a** = 4.023(1) Å and **c** = 8.719(4) Å for x = 0.10 phase. These lattice parameters are smaller than those reported for pure LaOFeAs (**a** = 4.038Å, **c** = 8.753Å) [14] due to smaller ionic size of $F^{-1}$ as compared to $O^{-2}$ ion. Fig.2 shows the energy dispersive X-ray spectra that confirms the presence of La, Fe As, O, F and Sb with La:Fe:As ~ 1:1:1. Small amount of Si is also indicated possibly due to the formation of SiO vapor from the quartz tube employed for synthesis.

The temperature dependence of resistivity at zero magnetic field is shown in Fig.2 for (a) $LaO_{0.8}F_{0.2}FeAs_{0.95}Sb_{0.05}$ and (b) $LaO_{0.8}F_{0.2}FeAs_{0.9}Sb_{0.10}$. The insets in figure 2a show the resistivity variation up to room temperature and the inductive part of the rf magnetic susceptibility attesting the onset of bulk diamagnetic state. With decreasing temperature, the resistivity decreases monotonously and a rapid drop to zero resistance state is observed starting at about 30.1 K and 28.6 K in samples *a* and *b* respectively. The criterion used for determination of transition temperature is shown schematically in the figure 2. The highest transition temperature for $LaO_{0.9}F_{0.1}FeAs$ is 28.5 K as reported earlier [11]. It is clear that the transition temperature is increased to 30.1 K for 5 % Sb substitution. The residual resistivity ratio (RRR = $\rho_{300}/\rho_{30}$) is 5.8 and 3.9 for $LaO_{0.8}F_{0.2}FeAs_{0.95}Sb_{0.05}$ and $LaO_{0.8}F_{0.2}FeAs_{0.9}Sb_{0.10}$ respectively. At room temperature, $\rho_{300 K}$ = 3.8 and 5.9 mΩ cm for samples *a* and *b* respectively which are of the same order as reported for La(O/F)FeAs [15]. Analyzing the behavior of resistivity in the normal state we observe that between 30 K and 150 K, the resistivity exhibits a dependence of the type $\rho = A + B T^2$ which implies that small angle scattering due to phonon modes has little impact on the observed resistivity (dominance of Umklapp processes). The inset in Fig 2b shows the linear fit between 35 K to 150 K which yields a value of A = 1.51 mΩ cm and B = 8.2 ×10$^{-5}$ mΩ cm K$^{-2}$. These values follow to a close approximation the values obtained for La(O/F)FeAs and indicate prevalence of strong electron-electron correlation with Sb doping [15].

Figure 3 illustrates the broadening of the superconducting transition in the presence of the applied magnetic field. To obtain information about the upper critical field ($H_{c2}$) and flux pinning properties, we have studied the variation of the electrical resistivity under various magnetic fields up to 5 T (Fig. 3). As the field is increased, one can see that the superconducting onset temperature (~ upper critical field $H_{c2}$) shifts with magnetic field weakly, but the zero resistance temperature (irreversibility field H* in analogy with High $T_c$ superconductors) decreases sharply with significant broadening of the superconducting transition. Using a criterion of 90% and 10% of normal state resistivity ($\rho_n$), we can estimate the upper critical field

$H_{c2}(T)$ and $H^*(T)$ respectively. The resultant H – T phase diagram for 5% Sb doped sample is shown in the inset of Figure 3. Further, we have attempted to calculate the zero temperature upper critical field $H_{c2}(0)$ using the Werthamer-Helfand-Hohenberg (WHH) formula $H_{c2}(0) = -0.693T_c (dH_{c2}/dT)$ [16]. The slope of $dH_{c2}/dT$ is estimated from the $H_{c2}$ versus T plots and are -3.34 and -3.70 T/K for samples a and b respectively. Using the transition temperature of $T_c = 30.1$ K and 28.6 K, we find $H_{c2}(0) = 69$ T and 73 T for $LaO_{0.8}F_{0.2}FeAs_{0.95}Sb_{0.05}$ and $LaO_{0.8}F_{0.2}FeAs_{0.9}Sb_{01}$ respectively. These values are more than twice as compared to the values reported in phase pure polycrystalline La(O/F)FeAs using the same 90% criteria [15]. Thus, due to the partial substitution of La by Sb, both the transition temperature and upper critical field show significant enhancement. Using the value of $H_{c2}(0)$, the estimated mean field Ginzburg-Landau coherence length ($\xi_{GL} = (\Phi_0 / 2\pi H_{c2})^{1/2}$) for the Sb doped LaOFFeAs samples is given by ~ 22 Å. Here $\Phi_0$ is the the flux quantum that equals to $2.07 \times 10^{-7}$ G-cm$^2$.

To get further insight into the transport properties of this new superconductor, we obtained the Seebeck Coefficient determined from the temperature-dependent thermoelectric power measurement (Fig. 4). At room temperature, the coefficient is negative (S = - 25 μV/K) and shows a minimum at 100 K (S = - 52 μV/K). Both the magnitude and the temperature corresponding to the minimum are much lower compared to that reported for La(O/F)FeAs [15]. Using the Mott expression $S = \pi^2 k_B T (2eT_F)$ we get the Fermi energy $E_F (= k_B T_F) \sim 0.084$ eV which is an order of magnitude less than cuprate superconductors [17]. However it should be pointed out that in a multiband scenario, extracting carrier concentration from thermo-power data would require supporting Hall and specific heat measurements.

While a clear understanding of $T_c$ enhancement with Sb doping in place of As would require an elaborate band structure calculation, from the above date we can propose a qualitative model. In pure La(O/F)FeAs the states near the Fermi level are primarily dominated by Fe $d$ states with dilute mixing of As $p$ states. There have been some reports predicting possible superconductivity in Fe-Sb based systems based on density functional calculations [18]. Particularly it has been shown that LiFeSb has similar electronic and magnetic structure with regards to the As based superconductors. Most importantly, a similar Fermi surface yet stronger spin density wave correlation is predicted. Our results indicating enhancement of $T_c$ on doping 'Sb' reinforces the above view and further studies on antimony-rich oxypnictide superconductors appear to be an interesting area for further research.

In summary, we have successfully synthesized polycrystalline $LaO_{0.8}F_{0.2}FeAs_{1-x}Sb_x$ superconductors (x = 0.05 and 0.10) by the conventional solid state reaction method. Our results show that the critical transition temperature increases to 30.1 K at x = 0.05 which is the highest transition temperature in the lanthanum based oxypnictides. A zero temperature upper critical field of ~73 Tesla with Ginzburg Landau coherence length $\zeta$ ~ 22 Å are estimated. Our results clearly demonstrate that disorder in the FeAs layer can lead to increase in transition temperature in this multiband superconductor.

**Acknowledgement :**

AKG and SP thank DST, Govt. of India for financial support. JP and SJS thank CSIR and UGC, Govt. of India, respectively for fellowships

**Figure captions:**

**Figure: 1.** Powder X-ray diffraction patterns of (a) $LaO_{0.8}F_{0.2}FeAs_{0.95}Sb_{0.05}$ and (b) $LaO_{0.8}F_{0.2}FeAs_{0.9}Sb_{0.10}$ annealed at 1150 C (The impurity phase is LaOF (*)) and (c) EDX spectra of $LaO_{0.8}F_{0.2}FeAs_{0.95}Sb_{0.05}$.

**Figure: 2.** Temperature dependence of zero field resistivity ($\rho$) as a function of temperature for (a) $LaO_{0.8}F_{0.2}FeAs_{0.95}Sb_{0.05}$ and (b) $LaO_{0.8}F_{0.2}FeAs_{0.9}Sb_{0.10}$. Inset of (a) shows resistivity up to room temperature and the inductive part of susceptibility. Inset of (b) shows resistivity up to room temperature and a linear fit to $\rho = A + B T^2$ equation.

**Figure: 3.** Temperature dependence of resistivity under various magnetic fields of (a) $LaO_{0.8}F_{0.2}FeAs_{0.95}Sb_{0.05}$ and (b) $LaO_{0.8}F_{0.2}FeAs_{0.9}Sb_{0.10}$ superconductors. Inset of figure (a) shows temperature dependence of upper critical field (■) and irreversibility field (●) as a function of temperature and inset of figure b the variation of thermoelectric power as a function of temperature for $LaO_{0.8}F_{0.2}FeAs_{0.95}Sb_{0.05}$.

**Figure 1.**

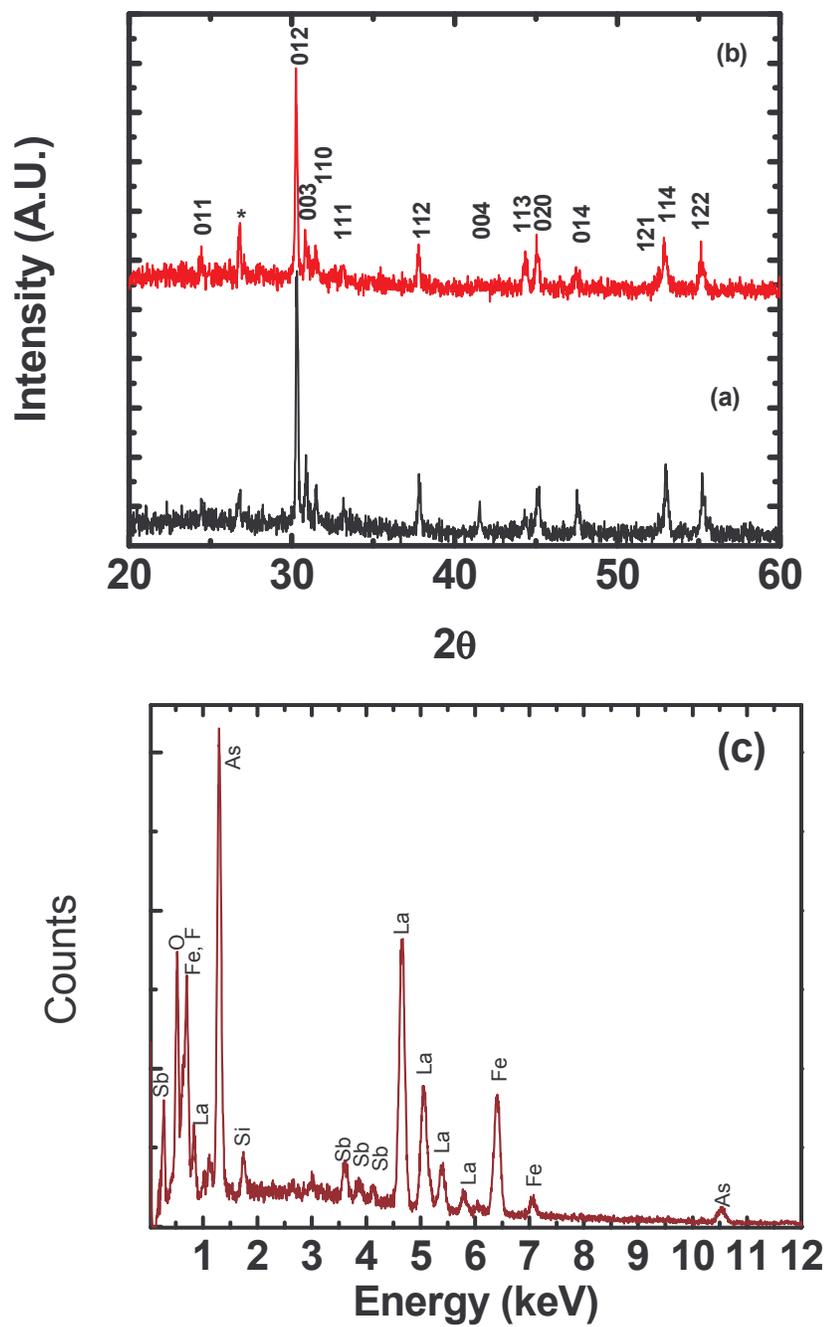

Figure 2.

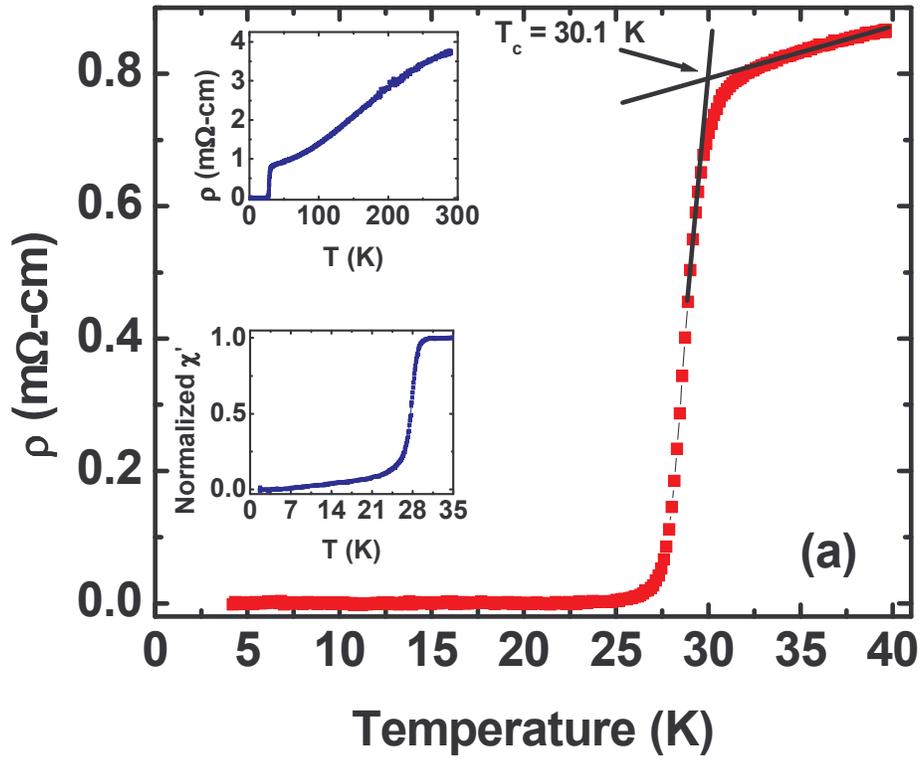

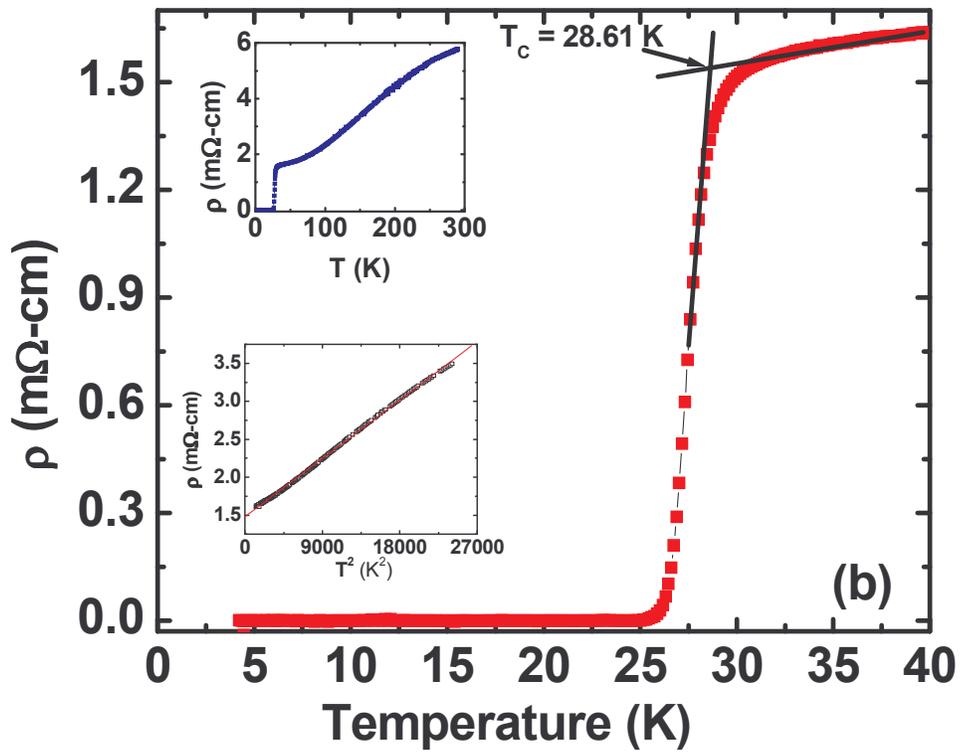

Figure 3

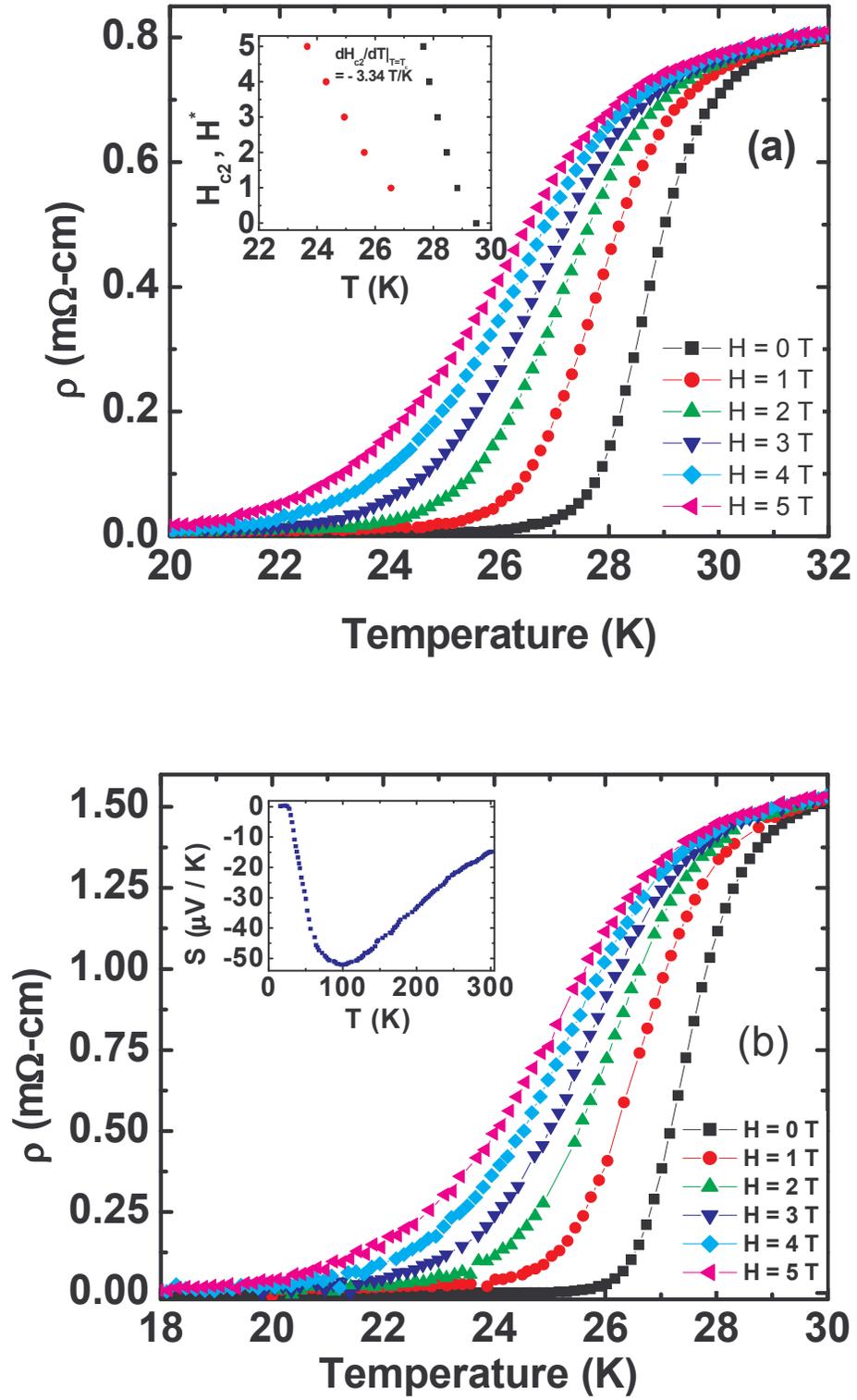